# SAXS/WAXS/DSC Study of Temperature Evolution in Nanopolymer Electrolyte

*Aleksandra Turković*[*,1], *Mario Rakić*[1], *Pavo Dubček1*, *Magdy Lučić-Lavčević*[2] *and Sigrid Bernstorff*[3]

[1]Institute "Ruđer Bošković", P.O. Box 180, HR-10002 Zagreb, Croatia

[2]Department of Physics, Faculty of Chemical Technology, University of Split, Teslina 10/V, 21000 Split, Croatia

[3]Sincrotrone Trieste, ss. 14, km 163,5 Basovizza, 34012 Trieste, Italy

[1]turkovic@irb.hr, [1]mrakic@irb.hr, [1]dubcek@irb.hr, [2]malula@ktf-split.hr

[3]Bernstorff@elettra.trieste.it

**ABSTRACT.** Electrolytes as nanostructured materials are very attractive for batteries or other types of electronic devices. $(PEO)_8ZnCl_2$ polymer electrolytes and nanocomposites $(PEO)_8ZnCl_2/TiO_2$ were prepared from PEO and $ZnCl_2$ and with addition of $TiO_2$ nanograins. The influence of $TiO_2$ nanograins was studied by small-angle X-ray scattering (SAXS) simultaneously recorded with wide-angle X-ray scattering (WAXS) and differential scanning calorimetry (DSC) at the synchrotron ELETTRA. It was shown by previous impedance spectroscopy (IS) that the room temperature conductivity of nanocomposite polymer electrolyte increased more than two times above 65°C, relative to pure composites of PEO and salts. The SAXS/DSC measurements yielded insight into the temperature-dependent changes of the grains of the electrolyte as well as to differences due to different heating and cooling rates. The crystal structure and temperatures of melting and crystallization of the nanosize grains was revealed by the simultaneous WAXS measurements.

## 1. Introduction

Understanding the structure of new materials on the mesoscopic scale (2-50 nm), such as clusters, aggregates and nanosized materials, requires suitable experimental techniques. Electromagnetic radiation can be used to obtain information about materials whose dimensions are on the same order of magnitude as the radiation wavelength. Scattering of X-rays is caused by differences in electron density. Since the larger the diffraction angle the smaller the length scale probed, wide angle X-ray scattering (WAXS) is used to determine the crystal structure on the atomic length scale while small-angle X-ray scattering (SAXS) is used to explore the microstructure on the nanometer scale.

SAXS experiments are suitable to determine the structure of nanocomposite polymer electrolyte. The solid electrolyte poly(ethylene oxide) (PEO) is one of the most extensively studied systems due to its relatively low melting point and glass transition temperature, $T_g$, its ability to play host to a variety of metal salt systems in a range of concentrations and recently it is material with the smallest ever polymer crystals prepared [1]. Polymeric complexes of $(PEO)_n$ with $ZnCl_2$ have been used, due to their stability and very high conductivity [2,3]. We have observed that the ionic conductivity at room temperature is up to two times larger compared to that above the phase transition temperature of 65°C [4].

The aim of the present investigation was to study the temperature behaviour of the $(PEO)_8ZnCl_2/TiO_2$ electrolyte by simultaneous SAXS/WAXS/DSC measurements. This structural investigation will provide answer to the question about the behaviour of nanosizes through the superionic phase transition, which occurs at ~65 °C.

## 2. Experimental

The polymer-salt complex was prepared by dissolving $ZnCl_2$ (p.a. Merck) and poly(ethylene oxide) (laboratory reagent, BDH Chemicals, Ltd., Poole, England, Polyox WSR-301, MW=$4\times10^6$, Prod 29740) in 50 % ethanol-water solution in stoichiometric proportions with addition of $TiO_2$ (Degussa P25) nanograins of average size of 25 nm and spherical shape [4].

Simultaneous SAXS/WAXD/DSC measurements were performed at the Austrian SAXS beamline at the synchrotron ELETTRA, Trieste [5]. Photon energy of 8 keV was used, and the size of the incident photon beam on the sample was 0.1 x 5 $mm^2$ (h x w). For each sample, SAXS and WAXS patterns were measured simultaneously in transmission setup using two 1D single photon counting gas detectors. The SAXS detector was mounted at a sample-to-detector distance of 1.75 m, corresponding to a q-range of 0.007 - 0.32 Å$^{-1}$. The WAXS detector was mounted to cover a d-spacing range of 0.32-0.94 nm.

The scattering wave vector, s equals s=2sinθ/λ=q/2π, where 2θ is the scattering angle and λ=0.154 nm the used wavelength. The method of interpreting the SAXS scattering data is based on the analysis of the scattering curve, which shows the dependence of the scattering intensity, I, on the scattering wave vector s.

The in-line micro-calorimeter built by the group of Michel Ollivon (CNRS, Paris, France) [6] was used to measure simultaneously SAXS/WAXS and high sensitivity DSC from the same sample. DSC phase transition temperature was determined at the intersection of tangent to the peak and the baseline.

SAXS is observed when electron density inhomogeneities of nanosized objects exist in the sample. If identical grains of constant electron density ρ are imbedded in a medium of constant ρ$_0$, only the difference Δρ=(ρ-ρ$_0$) will be relevant for scattering. If the grains are distanced from each other widely enough, it is assumed that they contribute to the scattered intensity independently. The central peak arises from all added secondary waves in phase at s=0. The amplitude is proportional to Δρ as only the contrast to the surrounding medium is effective. For the central part of the scattering curve, the universal Guinier approximation for all types of scattering objects/grains is valid [7-11]:

$$I_1(s) = \frac{1}{2\pi}(\Delta\rho)^2 \exp^{(-4\pi^2 s^2 R^2/3)} \tag{1}$$

where *R* is the gyration radius which is the average square distance from the centre of masses within the particles.

For WAXS the diameter of the nanocrystalline grains is obtained by the Debye-Scherer equation:

$$D = \frac{0.9 \cdot \lambda}{\beta \cdot \cos\theta} \qquad (2)$$

where λ is wavelength of the incident X-ray beam, and β is full width at half maximum (FWHM) of the WAXS line.

## 3. Results

Figure 1 shows the results from the SAXS measurements in the temperature range from 20°C to 100°C at rate of 1C°/min on polymer electrolyte $(PEO)_8ZnCl_2$ nanocomposite with $TiO_2$ nanograins which were performed at the SAXS-beamline of ELETTRA.

The intensity close to $I_s$ (for s=0) values shows changes throughout the phase transition at ~65°C. The intensity is lowering at 71°C indicating phase transition temperature in the heating cycle. After that intensity is gradually increasing till 53 °C. At this value the sudden drop shows the phase transition temperature in the cooling cycle. It is obvious that the hysteresis is present in this heating-cooling cycle.

Figure 2 shows the results from the SAXS, DSC and WAXS measurements on polymer electrolyte $(PEO)_8ZnCl_2$ nanocomposite with $TiO_2$ nanograins which were performed simultaneously at the SAXS-beamline of ELETTRA. The evolution of the average radii of grain sizes obtained by applying equation (1) is compared to the corresponding DSC and WAXS spectra behaviour. In the heating cycle the superionic phase transition can be seen as the sudden drop of the nanograin sizes at the phase transition temperature. In the cooling part a hysteresis can be seen as the phase transition occurs at lower temperature. The endothermic and exothermic peaks found in DSC during the same temperature cycle are in agreement with the sudden changes in the average nanograin sizes as obtained from the SAXS measurements and drops of the intensity in the WAXS spectra. In the heating cycle with rate of 1°C/min in the SAXS data there are two trends, first an increasing of the grain size up to 71°C and then a sudden drop at this phase transition temperature.

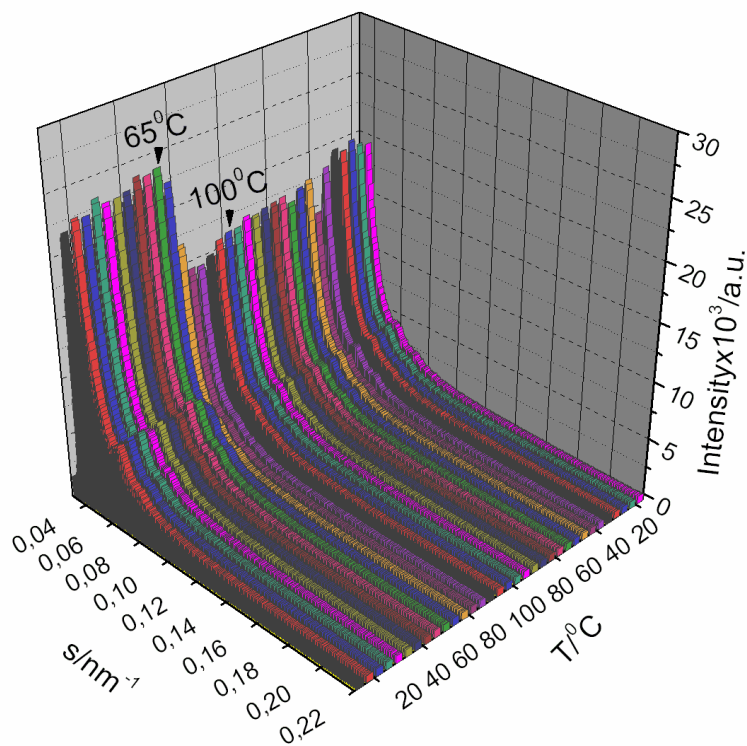

*Figure 1. The results from the SAXS measurements in the temperature range from 20°C to 100°C at rate of 1C°/min on polymer electrolyte (PEO)$_8$ZnCl$_2$ nanocomposite with TiO$_2$ .*

The average radius of grains varies from 6.1 nm to 6.4 nm in the region below the phase transition temperature and then from 6.0 nm to 6.1 nm in the highly conductive phase of the polymer electrolyte. The DSC spectrum shows the start of the endothermic peak at 68°C in the heating cycle. The cooling cycle in the SAXS data shows the start of the transition at 53ºC, and the change of grain sizes in the range from 6.2 nm to 6.0 nm. The WAXS spectra show drop of the intensity at 65°C and in cooling cycle the temperature of the start of recrystallization is 51 °C. In SAXS measurements the second run with the rate of 1°C/min (Fig. 3), results in changes of grain sizes from 5.8 nm to 6.3 nm, the third run with 3°C/min (Fig. 4), registers changes from 6.0 nm to 6.3 nm and during the forth run with 5°C/min (Fig. 5), grain sizes change from 6.1 nm to 6.2 nm. From these we can generally conclude that average grain sizes in all four runs remained in the same range from 5.8 nm to 6.3 nm. In a lamellar picture of PEO [12, 13] these sizes of grains would correspond to the lamellae LP2 with no integrally folded (NIF) chains [14, 15] combined with salt and $TiO_2$.

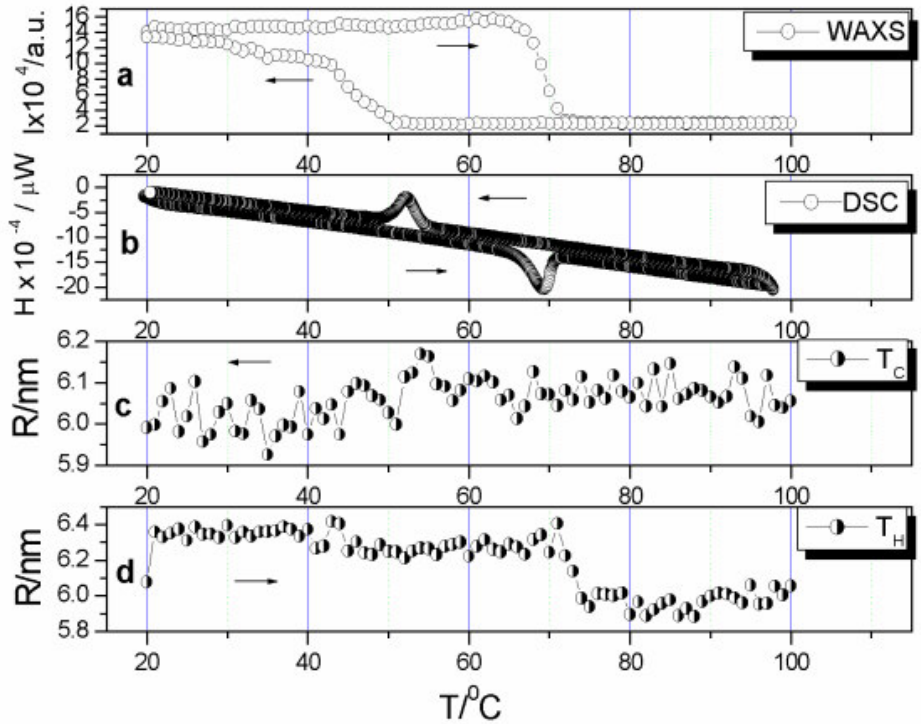

*Figure 2. SAXS, DSC and WAXS results for nanocomposite polyelectrolyte $(PEO)_8ZnCl_2/TiO_2$ in the temperature range from 20°C to 100°C at rate of 1C°/min, a) showing evolution of the intensity of the strong WAXS line at $2\theta =19.21$ (denoted with WAXS), b) showing the DSC heating and cooling cycle (denoted with DSC), c) showing radius of gyration ($R_G$) in the cooling cycle of SAXS (denoted with $T_C$) and d) presenting $R_G$ in the heating cycle of SAXS (denoted with $T_H$).*

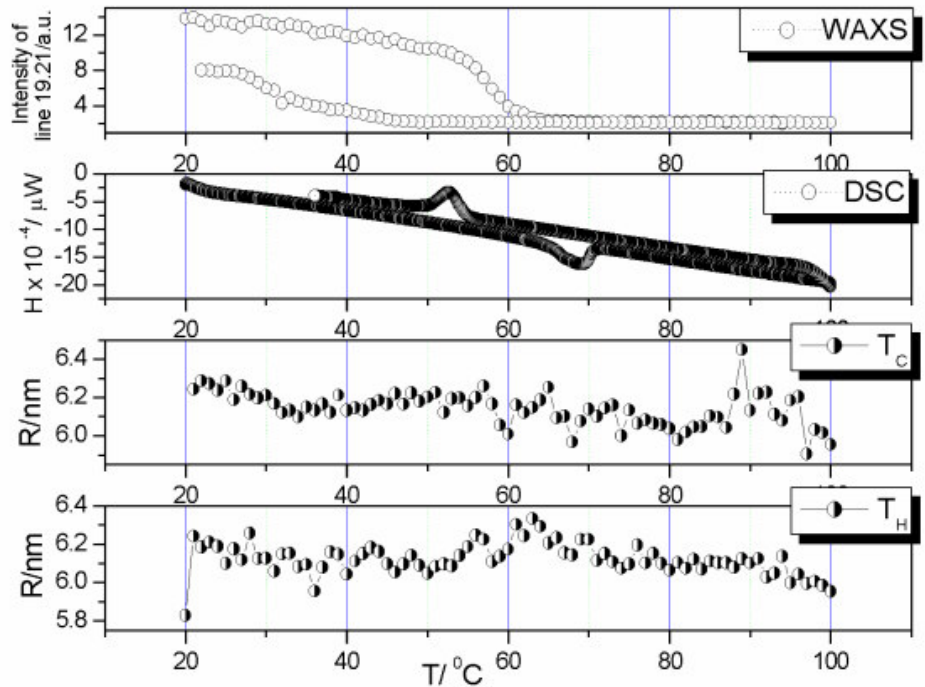

*Figure 3. WAXS, DSC and SAXS results for nanocomposite polyelectrolyte $(PEO)_8ZnCl_2/TiO_2$ in the temperature range from 20°C to 100°C at rate of 1C°/min second time, a) showing evolution of the intensity of the strong WAXS line at $2\theta = 19.21$ (denoted with WAXS), b) showing the DSC heating and cooling cycle (denoted with DSC), c) showing radius of gyration ($R_G$) in the cooling cycle of SAXS (denoted with $T_C$) and d) presenting $R_G$ in the heating cycle of SAXS (denoted with $T_H$).*

The DSC spectrum for the rate of 1°C/min shows that the phase transition temperature is 55°C, which is determined at the beginning of the peak in the cooling cycle. Both, the SAXS and DSC data show a hysteresis, i.e. much lower phase transition temperatures than 65°C in the cooling cycle. This temperature is the melting temperature of the PEO crystallites i.e. "spherulites" [16-18]. In the case of the nanocomposite polymer electrolyte, combined forms of PEO and $ZnCl_2$, $ZnCl_2$ or both in combination with $TiO_2$ crystallites, influence the melting temperature. The combined WAXS, SAXS and DCS results are presented in the Table 1.

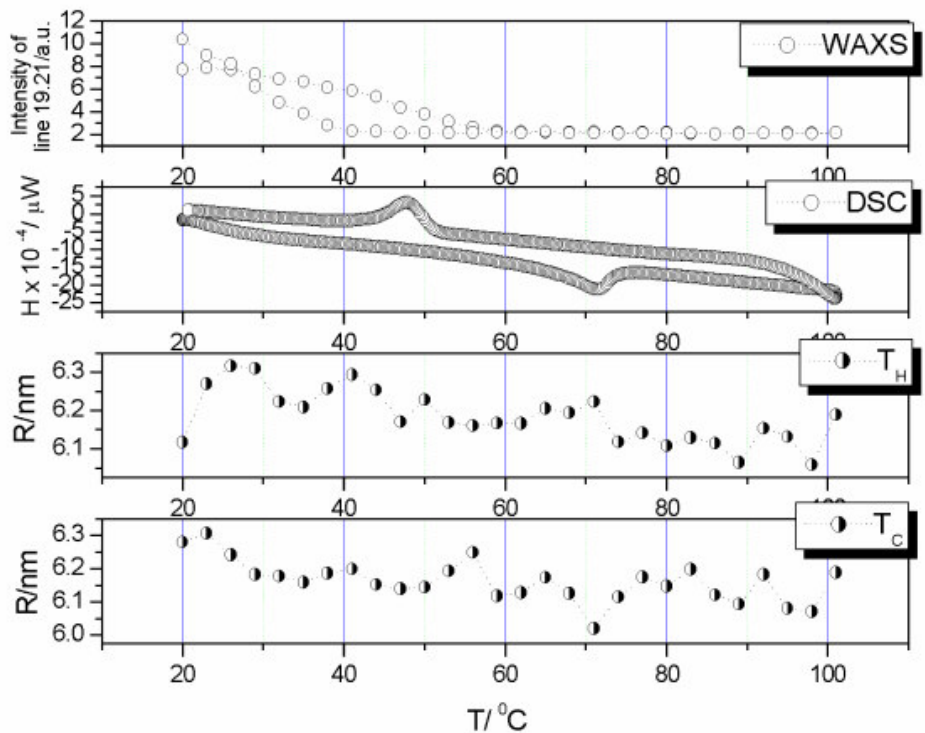

*Figure 4. WAXS, DSC and SAXS results for nanocomposite polyelectrolyte $(PEO)_8ZnCl_2/TiO_2$ in the temperature range from 20°C to 100°C at rate of 3C°/min second time, a) showing evolution of the intensity of the strong WAXS line at $2\theta =19.21$ (denoted with WAXS), b) showing the DSC heating and cooling cycle (denoted with DSC), c) showing radius of gyration ($R_G$) in the cooling cycle of SAXS (denoted with $T_C$) and d) presenting $R_G$ in the heating cycle of SAXS (denoted with $T_H$).*

The WAXS recordings were done simultaneously with the SAXS and DSC measurements. In Figure 6a) and b) the WAXS results are shown as 3D plots for a heating and a cooling cycle at rate of 1°C/min. In the heating cycle the phase transition occurs at 72 °C and in the cooling cycle there is again a hysteresis and the recrystallization occurs at 53 °C. Above the phase transition temperature the WAXS spectra are registering the amorphous phase of the polymer electrolyte. Accordingly, the assignation of these WAXS lines [19] in the heating and the cooling cycle with the rate of 1°C/min is given in the Table 2.

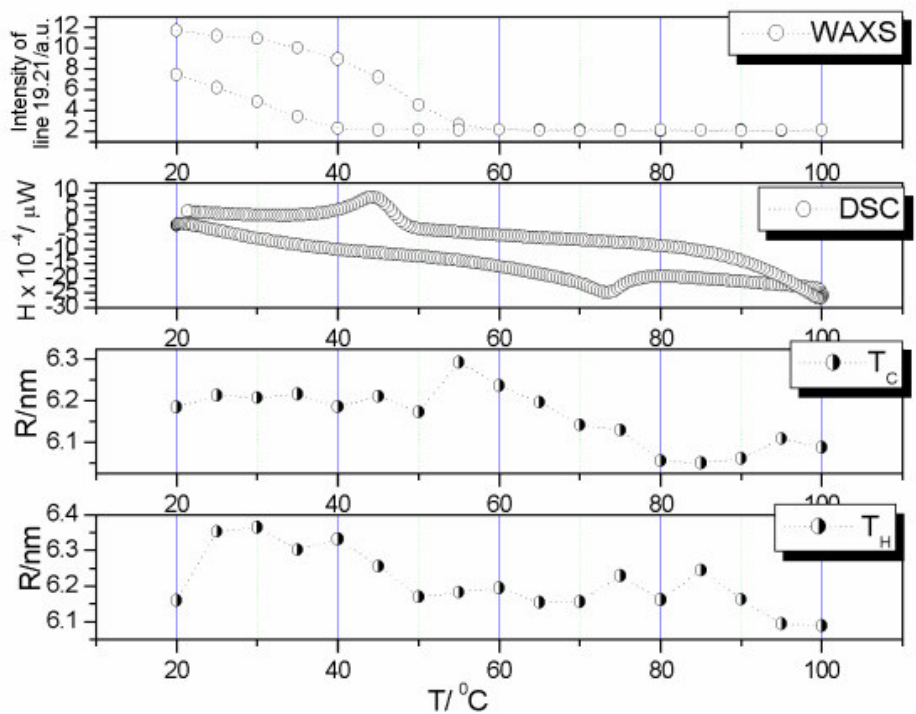

*Figure 5. WAXS, DSC and SAXS results for nanocomposite polyelectrolyte $(PEO)_8ZnCl_2/TiO_2$ in the temperature range from 20°C to 100°C at rate of 5C°/min second time, a) showing evolution of the intensity of the strong WAXS line at $2\theta =19.21$ (denoted with WAXS), b) showing the DSC heating and cooling cycle (denoted with DSC), c) showing radius of gyration ($R_G$) in the cooling cycle of SAXS (denoted with $T_C$) and d) presenting $R_G$ in the heating cycle of SAXS (denoted with $T_H$).*

The columns in the Table 2 for 72°C and 53°C are shown with shaded cells as these are the WAXS spectra with amorphous halo and no crystalline lines are present.

The most intensive lines in WAXS attributed to combined $PEO/ZnCl_2$ crystallites disappear at 72°C, 65°C, 56°C and 55 °C for the heating rates of 1°C /min, 1°C /min, 3°C /min and 5°C/min respectively, but lines of lower intensity representing the PEO "spherulites" are disappearing at somewhat lower temperatures. In the cooling cycle the strongest WAXS lines re-appear at 53°C, 46°C, 38 °C and 35 °C and the lines of lower intensities re-appear at lower temperatures.

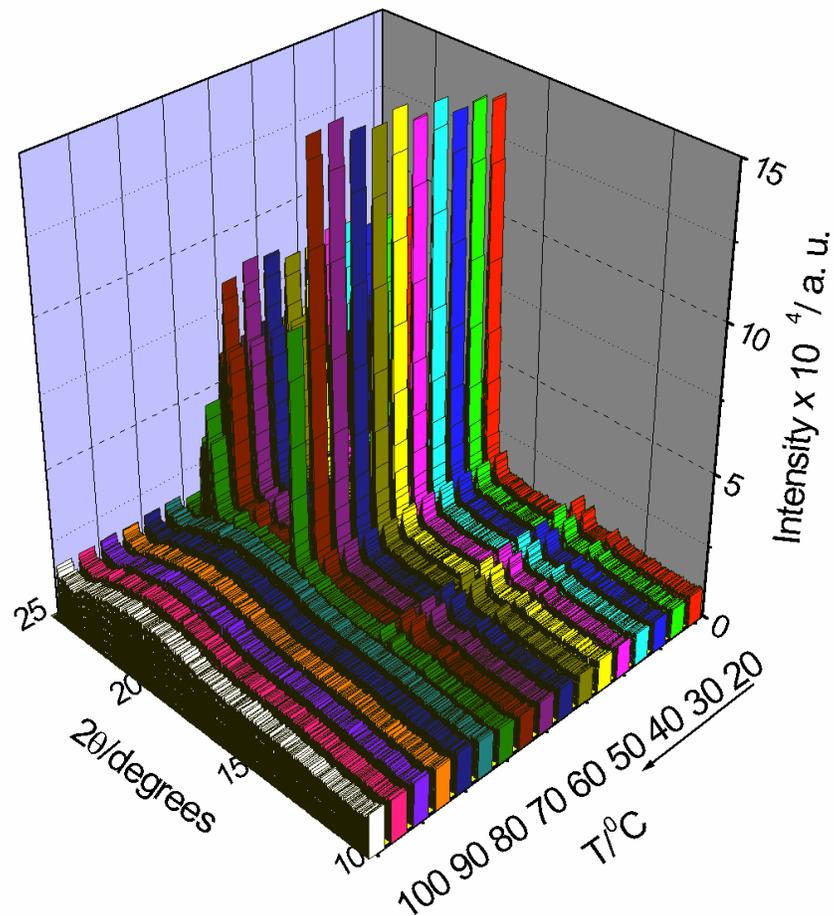

*Figure 6. a) 3D WAXS of $(PEO)_8ZnCl_2/TiO_2$ with a heating rate of 1°C/min.*

The WAXS obtained radius values for the grain sizes are higher than the SAXS values. For the rate of 1°C/min WAXS values are equal to 27.3 nm at 20°C, increasing up to 43.3 nm at 72°C. In the cooling cycle the grain size is 23.2 nm at 53°C, and again is 37.4 nm at 20°C. The WAXS data are giving the information of the side chain order. During the heating and cooling cycles the side chain ordering is giving us information of the lateral domain sizes or "spherulites".

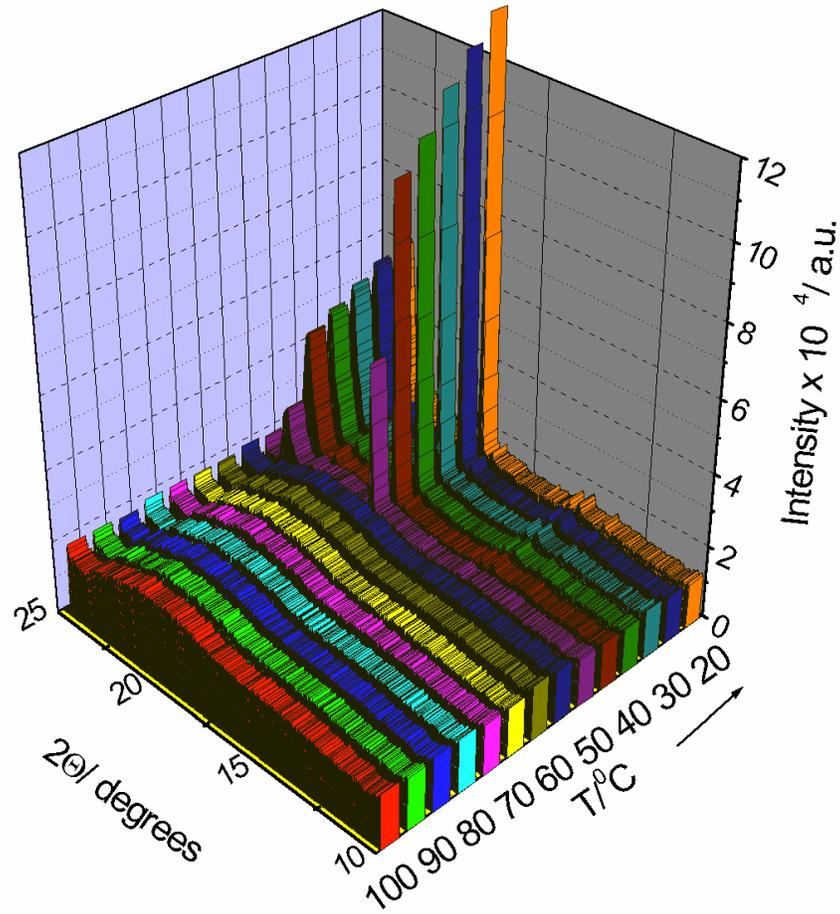

*Figure 6. b) 3D WAXS of (PEO)$_8$ZnCl$_2$/TiO$_2$ with a cooling rate of 1°C/min.*

In the heating cycles the lateral domain sizes are increasing till the phase transition in the cases with the rate 1°C /min, decreasing for 3°C /min and staying the same for 5°C/min. In the cooling cycle the lateral domain sizes are showing trend of returning close to the starting room temperature size in the cases with the rate of 1°C/min. The case of 3°C/min is showing decrease of domain size and the rate with 5°C /min stays the same with small change of domain size. The WAXS results for all heating and cooling rates are, together with SAXS and DSC data, presented in the Table 1.

**Table 1.** Changes of average grain radius <R>/ nm calculated by (1), R=D/2 as determined from (2) and phase transition temperatures t (in °C) in $(PEO)_8ZnCl_2/TiO2$ polyelectrolyte during heating and cooling as determined by SAXS/WAXS/DSC measurements.

| $(PEO)_8ZnCl_2$/ $TiO_2$ | heating | | | | |
| --- | --- | --- | --- | --- | --- |
| | **SAXS** | | **WAXS** | | **DSC** |
| heating *(°C/min)* | *t (°C)* | *R (nm)* | *t (°C)* | *R (nm)* | *t (°C)* |
| 1 | 71 | 6.1-6.4 <br> 6.0-6.1 | 71 | 27.3-43.3 | 66 |
| 1 | 63 | 5.8-6.3 <br> 6.1-6.0 | 65 | 30.5-41.6 | 62 |
| 3 | 65 | 6.3-6.2 <br> 6.0-6.2 | 56 | 34.6-27.5 | 61 |
| 5 | / | 6.2-6.1 | 55 | 36.3-35.9 | 60 |
| $(PEO)_8ZnCl_2$/ $TiO_2$ | cooling | | | | |
| | **SAXS** | | **WAXS** | | **DSC** |
| cooling *(°C/min)* | *t (°C)* | *R (nm)* | *t (°C)* | *R (nm)* | *t (°C)* |
| 1 | 53 | 6.1-6.4 <br> 6.1-6.0 | 53 | 23.2-37.4 | 55 |
| 1 | 57 | 6.0-6.3 <br> 6.1-6.2 | 46 | 33.0-36.3 | 54 |
| 3 | 50 | 6.2-6.2 <br> 6.2-6.1 | 38 | 30.8-43.3 | 52 |
| 5 | / | 6.1-6.2 | 35 | 36.4-36.9 | 49 |

The differences in the phase transition temperatures for different heating and cooling rates for all SAXS, WAXS and DSC results are evident. This shows the importance of use the data of the same experimental conditions in the cases of comparing the literature results. Also, the differences in the values of the phase transition temperatures are obvious in-between all three methods. The best agreement between phase transition temperatures among the four runs is achieved for the first run with heating-cooling rate of 1°C/min. At the second run with 1°C/min and the two higher rates of 3°C/min and 5°C/min the differences in the phase transition temperature values are increasing.

**Table 2.** Assignments of the (PEO)$_8$ZnCl$_2$/TiO$_2$ WAXS lines for the run with the rate of 1°C/min according to values reported in the Powder Diffraction File PDF-2 (2003) cards numbers: 16-0869 and 49-2201 [18].

| Assignation | (PEO)$_8$Zn Cl$_2$ | | | | | | | |
|---|---|---|---|---|---|---|---|---|
| | heating | | | | cooling | | | |
| | 20°C | | 71°C | 72°C | 53°C | 52°C | 20 °C | |
| | 2θ(deg) | Int / h k l | 2θ(deg) | 2θ(deg) | 2θ(deg) | 2θ(deg) | 2θ(deg) | Int/h k l |
| 1=PEO | 13.17 | 6 | | | | | / | / |
| 2=PEO | 13.57 | 5 | | | | | 13.59 | 5 |
| 3=PEO | 14.69 | 9 | | | | | 14.67 | 9 |
| 4=PEO | 15.12 | 7 | | | | | 15.14 | 7 |
| 5=PEO/TiO$_2$ | 18.66 | 9 | | | | | 18.66 | 9 |
| 6=PEO/ZnCl2 | 19.21 | 75/ 0 2 1 | 19.13 | | | 19.14 | 19.23 | 75/ 0 2 1 |
| 7=ZnCl$_2$ | 21.15 | 20/ $\bar{1}$ 1 2 | | | | | 21.13 | 20/ $\bar{1}$ 1 2 |
| 8=PEO | 21.54 | 19 | | | | | / | / |
| 9=ZnCl$_2$ | 22.07 | 50/ 1 2 1 | | | | | 22.07 | 50/ 1 2 1 |
| 10=PEO | 22.67 | 100 | 22.96 | | | | / | / |
| 11= PEO/ZnCl2 | 23.05 | 100/ 0 1 3 | | | | | / | / |
| 12= PEO/ZnCl2 | 23.29 | 100/ 0 1 3 | 23.22 | | | | 23.42 | 100/01 3 |
| 13=ZnCl$_2$ | 23.59 | 10/ 0 3 1 | | | | | / | / |

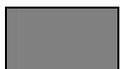 Denotes amorphous phase/halo in the material

It leads to the conclusion about the necessity to achieve a good equilibrium conditions in order to obtain agreement between values for the phase transition temperatures in all three methods. Slower rates than rates of 1°C /min should be tested in order to achieve better agreement between results of WAXS, DSC and SAXS measurements for this system.

The combination of the three methods reveals the nature of the physical transformation of the polymer electrolyte into a super ionic conductor. The nanocomposite crystalline and amorphous polymer matrix is turning into an amorphous highly conductive phase. In contrast to WAXS, which exhibits lines and crystalline grains only in the low temperatures crystalline phases, SAXS is showing the existence of nanograins in both the low and high temperature phase. At the phase transition temperature the grain size changes, it is becoming smaller at higher temperatures. The nature of the nanograins as seen by SAXS is not just the pure crystalline, but also the partly amorphous form, while WAXS records only pure crystalline nanograins. Thus the picture of the highly conductive phase consists of a completely amorphous polymer matrix, which is known to be suitable for ion–conduction by elastic movement of PEO chains, and of nanograins of combined PEO/$ZnCl_2$ and $TiO_2$ structures, which could also contribute to $Zn^{2+}$-ion conduction by a hopping mechanism. Under proper circumstances, the presence of ion-transport pathways can be as important as polymer segmental motion [20-22].

The same combined SAXS/WAXS/DSC experiments were also performed on polymer electrolytes treated by irradiation with γ-rays and by simultaneously adding of $TiO_2$ nanograins of 25 nm size. These results will be presented in the next paper.

4. Conclusion

The combined SAXS/WAXS/DSC measurements have shown that the nanostructure of the polymer electrolyte (PEO)$_8$ZnCl$_2$/TiO$_2$ is changing during the crystalline-amorphous phase transition to a highly conductive superionic phase. The significant role that the nanodimensions of the electrolyte material play in the Zn2+-ion mobility was discussed. The combined SAXS/WAXS information about the evolution of the average grain sizes during the phase transition gave insight into the nanomorphology, which influences the ionic transport in a nanocomposite polymer electrolyte. Further optimizations of the electrolyte properties are in progress since these nanostructured materials are very attractive for batteries or other types of electronic devices.

Acknowledgment

The Ministry of Education, Science and Sport of the Republic of Croatia is thanked for support of this work.